\RequirePackage[table]{xcolor}
\documentclass[journal,twoside,web]{ieeecolor}
\let\labelindent\relax
\usepackage{generic}
\usepackage{amsmath,amssymb,amsfonts}
\usepackage{algorithmic}
\usepackage{graphicx}
\usepackage{algorithm}
\usepackage{textcomp}
\usepackage{svg}
\usepackage{tablefootnote}
\usepackage[font=small,labelfont=bf]{caption}
\usepackage{subcaption}
\usepackage{multirow}
\usepackage{booktabs}
\usepackage{tabularx}
\usepackage{float}

\usepackage{cite}
\usepackage{enumitem}
\usepackage{orcidlink}

\usepackage{hyperref}
\hypersetup{
    colorlinks=true,       
    linkcolor=blue,        
    citecolor=green,       
    filecolor=magenta,     
    urlcolor=cyan,         
}

\def\BibTeX{{\rm B\kern-.05em{\sc i\kern-.025em b}\kern-.08em
    T\kern-.1667em\lower.7ex\hbox{E}\kern-.125emX}}
\begin{document}
\title{S\textsuperscript{3}F-Net: A Multi-Modal Approach to Medical Image Classification via Spatial-Spectral Summarizer Fusion Network}

\author{Md. Saiful Bari Siddiqui~\orcidlink{0009-0000-7781-0966},~\IEEEmembership{Graduate Student Member,~IEEE},
Mohammed Imamul Hassan Bhuiyan~\orcidlink{0009-0009-6767-9357},~\IEEEmembership{Senior Member, IEEE.}
\thanks{Received 2 September 2025; Revised 27 November 2025: Accepted 3 April 2026. (Corresponding author: Md. Saiful Bari Siddiqui).}
\thanks{Md. Saiful Bari Siddiqui is with the Department of Computer Science and Engineering, BRAC University, Dhaka, Bangladesh (e-mail: saiful.bari@bracu.ac.bd).}
\thanks{Mohammed Imamul Hassan Bhuiyan is with the Department of Electrical and Electronic Engineering, Bangladesh University of Engineering and Technology, Dhaka, Bangladesh (e-mail: imamul@eee.buet.ac.bd).}}

\maketitle

\begin{abstract}
Convolutional Neural Networks (CNNs) have become a cornerstone of medical image analysis due to their proficiency in learning hierarchical spatial features. However, this focus on a single domain is inefficient at capturing global, holistic patterns and fails to explicitly model an image's frequency-domain characteristics. To address these challenges, we propose the Spatial-Spectral Summarizer Fusion Network (S\textsuperscript{3}F-Net), a dual-branch framework that learns from both spatial and spectral representations simultaneously. The S\textsuperscript{3}F-Net performs a fusion of a deep spatial CNN with our proposed shallow spectral encoder, SpectraNet. SpectraNet features the proposed SpectralFilter layer, which leverages the Convolution Theorem by applying a bank of learnable filters directly to an image's full Fourier spectrum via a computation-efficient element-wise multiplication. This allows the SpectralFilter layer to attain a global receptive field instantaneously, with its output being distilled by a lightweight summarizer network. We evaluate S\textsuperscript{3}F-Net across four diverse medical imaging datasets spanning different scales and modalities: HAM10000 (dermoscopy), BUSI (ultrasound), BRISC2025 (MRI), and Chest X-Ray Pneumonia (radiography), to validate its efficacy and generalizability, and reveal the task-dependent nature of the optimal fusion strategy. Our framework consistently and significantly outperforms its strong spatial-only baseline in all cases, with accuracy improvements of up to \textbf{5.13\%}. With a powerful Bilinear Fusion, S\textsuperscript{3}F-Net achieves a state-of-the-art competitive accuracy of 98.76\% on the BRISC2025 dataset. A simpler Concatenation Fusion performs better on the texture-dominant Chest X-Ray Pneumonia dataset, achieving 93.11\% accuracy, surpassing many top-performing, much deeper models. Our explainability analysis also reveals that the S\textsuperscript{3}F-Net learns to dynamically adjust its reliance on each branch based on the input pathology. These results verify that our dual-domain approach is a powerful and generalizable paradigm for medical image analysis. Our source codes and pre-trained model weights are publicly available on GitHub at: \url{https://github.com/Saiful185/S3F-Net}.
\end{abstract}


\section{Introduction} \label{intro}
The application of deep learning, particularly Convolutional Neural Networks (CNNs), has led to transformative advances in the field of automated medical image analysis. Architectures such as ResNet and VGG have demonstrated remarkable success in a variety of classification tasks~\cite{simonyan2014vgg},~\cite{szegedy2016inceptionv3},~\cite{he2016resnet}. The core strength of these models lies in their ability to learn a rich hierarchy of spatial features like edges and textures, progressively building an understanding of complex structures.

However, the operating principle of standard CNNs, which relies on stacked layers of local convolutional kernels, is fundamentally biased towards spatial feature extraction. Although powerful, this approach presents two key limitations that are particularly crucial in the biomedical image analysis domain. Firstly, CNNs are inherently local operators. To capture global, holistic patterns, they must progressively build a large receptive field through significant network depth. This is particularly challenging for high-resolution medical images, where even deep networks may struggle to efficiently integrate long-range dependencies, such as the diffuse textural changes indicative of certain pathologies. A neuron in a deep layer must aggregate information from a long chain of preceding local operations. This hierarchical process is an indirect and computationally inefficient method for perceiving global context. Moreover, the fundamental architecture of a CNN does not explicitly model the frequency-domain characteristics of an image. These spectral properties, describing the periodic and textural nature of an image, can hold crucial diagnostic information but are often lost or represented inefficiently by a purely spatial network. For instance, diagnosing a skin lesion requires analyzing both its border shape (a local, spatial feature) and its internal color variegation (a global, textural feature better captured in the spectral domain). Similarly, the diffuse, low-frequency textural changes indicative of pneumonia in radiography are challenging for the limited receptive fields of typical convolutional kernels. Modalities like ultrasound and radiography are rich in these textural patterns, providing a high-fidelity signal for frequency analysis that is largely overlooked by CNNs.

Historically, attempts to integrate frequency-domain analysis into neural networks have faced their own challenges. Early approaches often focused on using the Fast Fourier Transform (FFT) as a tool for accelerating convolutions~\cite{vasilache2015fast},~\cite{mathieu2013fast}, rather than as a feature extraction mechanism itself. The prospect of using learnable filters directly in the frequency domain was largely hindered by the fact that representing a filter at full spectral resolution required an excessive number of parameters, creating a high risk of severe overfitting, especially on typically smaller medical datasets. However, the core operation of CNNs, spatial convolution, is computationally intensive for both the forward and backward passes. Although the Convolution Theorem states that this expensive convolution in the spatial domain is equivalent to a simple element-wise multiplication in the frequency domain, an operation with a much lower computational complexity of $O(N\log N)$ via the Fast Fourier Transform (FFT) versus $O(N^2)$ for spatial convolution~\cite{chi2020fast}, harnessing this efficiency in a learnable feature-extraction context remained a challenge. 

To bridge these gaps, we introduce the \textbf{Spatial-Spectral Summarizer Fusion Network (S\textsuperscript{3}F-Net)}, a novel dual-branch framework that explicitly and efficiently leverages both spatial and spectral information. Our central hypothesis is that by fusing features from these two complementary domains, our model can develop a more robust and comprehensive understanding of medical images. The S\textsuperscript{3}F-Net is comprised of two specialized, asymmetric towers. The first is a deep, standard CNN that acts as a powerful morphological feature extractor. The second, a key contribution of this work, is our novel \textbf{SpectraNet} branch. Our work overcomes the historical challenges of learnable spectral models through two key architectural innovations. First is the \textbf{SpectralFilter Layer}, which applies a bank of learnable filters directly to an image's entire Fourier spectrum via a parameter-efficient element-wise multiplication, attaining a global receptive field from the very first layer. Second, we introduce the \textbf{SpectraNet} branch, a deliberately shallow architecture built upon this layer. \textbf{We posit that since a global view is immediately available in the frequency domain, a deep hierarchy is unnecessary.} Instead, the \textbf{shallow} SpectraNet branch focuses on summarizing this global information, distilling it through a lightweight ``funnel'' head comprised of Depthwise Separable Convolutions and dense layers to produce a small but powerful feature vector.


A central finding of this research is that the optimal strategy for combining the information from the two domains is highly task-dependent. We systematically investigate two primary fusion mechanisms at the vector level. The first is a simple \textbf{Concatenation Fusion}, which appends the two feature vectors and allows a subsequent dense layer to learn their relationships. The second is a more powerful \textbf{Bilinear Fusion}, which computes the outer product of the two vectors. This mechanism explicitly models the rich, pairwise interactions between every spatial feature and every spectral feature, making it suitable for complex tasks where the interplay between morphology and texture is critical.

The contributions of this paper are as follows:
\begin{itemize}
    \item We introduce the \textbf{SpectralFilter Layer}, a computationally efficient deep learning component that applies learnable filters directly to an image's full Fourier spectrum via element-wise multiplication.
    \item We present the \textbf{S\textsuperscript{3}F-Net}, a flexible dual-domain framework featuring our proposed shallow spectral encoder, \textbf{SpectraNet}, that outperforms its strong spatial-only baseline across four diverse medical imaging modalities.
    \item We demonstrate that the optimal fusion strategy for the S\textsuperscript{3}F-Net is modality and task-dependent.
    \item We present SOTA competitive results on multiple benchmarks, achieving \textbf{98.76\%} accuracy on BRISC2025 and \textbf{93.11\%} accuracy on the Chest X-Ray Pneumonia dataset.
    \item We develop a custom explainability metric to quantify branch contributions and reveal S\textsuperscript{3}F-Net's ability to dynamically weigh each branch based on the pathology.
\end{itemize}

\section{Related Works} \label{related_work}


\subsection{Deep Learning in Medical Image Analysis}
Convolutional Neural Networks (CNNs) have become the de facto standard for a wide array of medical image analysis tasks. Landmark architectures such as VGGNet~\cite{simonyan2014vgg}, Inception-v3~\cite{szegedy2016inceptionv3}, ResNet~\cite{he2016resnet}, DenseNet~\cite{huang2017densenet}, MobileNet~\cite{howard2017mobilenets}, and EfficientNet~\cite{tan2019efficientnet} have established the power of deep networks for feature extraction. In the medical domain, these models, typically pre-trained on large-scale datasets like ImageNet~\cite{deng2009imagenet}, are fine-tuned to achieve state-of-the-art results. These architectures have been successfully applied to diverse tasks, including classifying skin lesions~\cite{khan2021skin}, detecting pneumonia~\cite{Alshanketi2025Pneumonia}, etc. 

To overcome the inherent locality of standard convolutions, recent research has focused on integrating more sophisticated attention mechanisms. For complex inputs like 3D volumes or multi-phase scans, methods such as the CdTransformer~\cite{Zhu2024transformer} and multi-phase self-attention~\cite{Xu2023Liver} have been developed to better model long-range spatial and temporal dependencies.

While these advanced methods refined the feature extraction process \textit{within the spatial domain}, our work explores a fundamentally different and complementary direction. We posit that a truly holistic understanding requires simultaneous learning from a complementary mode of signal representation: the frequency domain.

\subsection{Frequency Domain in Deep Learning}
The use of the Fourier Transform in neural networks has a rich history. Early works by Mathieu et al.~\cite{mathieu2013fast} and Vasilache et al.~\cite{vasilache2015fast} provided a theoretical and practical foundation, demonstrating that the Convolution Theorem could be leveraged via FFT to accelerate CNNs training on GPUs.

A conceptual shift began with works that explored the representative power of the spectral domain itself. Rippel et al. first proposed the use of spectral representations for weight parameterization and later introduced Spectral Pooling, which performs downsampling by cropping the low-frequency components of the Fourier spectrum, providing a theoretically sound, anti-aliasing alternative to max-pooling~\cite{rippel2015spectral}. Pratt et al. followed with the Fourier Convolutional Neural Network (FCNN), one of the first architectures to learn filters directly in the frequency domain, demonstrating its viability~\cite{pratt2017fcnn}. Li et al. shed light on the optimization landscape in their recent theoretical analysis of Spectral Neural Networks~\cite{li2023spectral}. However, these early, "pure" spectral approaches were often limited in practice by a critical weakness: representing a learnable filter at the full resolution of the Fourier spectrum led to a massive number of parameters, creating a high risk of severe overfitting, particularly on datasets smaller than ImageNet. Our S\textsuperscript{3}F-Net directly addresses this challenge through its asymmetric design and parameter-efficient summarizer head.


Beyond the Fourier Transform, other spectral representations like the Discrete Cosine Transform (DCT) and the Discrete Wavelet Transform (DWT) have been integrated into deep learning frameworks. The DWT, in particular, has proven effective in capturing multi-scale spatial-frequency information. Recent works have explored hybrid approaches, such as combining Wavelet and Fourier convolutions for ultrasound reconstruction~\cite{jeong2025r2b}, demonstrating the power of leveraging multiple transform domains. Wavelets have also been explored for super-resolution~\cite{zhong2018joint}. S\textsuperscript{3}F-Net differs fundamentally from wavelet-based approaches in its specific focus on the \textbf{global context} provided by the Fourier domain. While wavelets excel at a multi-scale, localized analysis, our \textbf{SpectralFilter Layer} is designed to capture the non-local textural and periodic properties of the entire image in a single, efficient operation.

A notable advancement is the introduction of attention mechanisms in the frequency domain. FcaNet~\cite{qin2021fcanet}, for example, proposes a frequency channel attention mechanism. This method compresses each feature channel's spatial information and uses a 1D DCT to analyze the frequency content \textit{along the channel axis}, allowing the network to selectively weight different feature channels. While FcaNet uses the frequency domain to generate an attention vector for re-weighting existing spatial channels, our approach is fundamentally different. Our \textbf{SpectralFilter Layer} operates on the \textbf{2D spatial frequency spectrum} of each channel via a 2D FFT, allowing it to learn filters that are sensitive to spatial textures and periodic patterns directly, thereby generating new, explicitly spectral features.

The Fast Fourier Convolution (FFC) proposed by Chi et al.~\cite{chi2020fast} also leverages learnable filters in the frequency domain. FFC's design relies on a layer-level channel split, where a fraction of the cropped feature channels are processed by its global spectral path. This approach, while computationally efficient, means that neither the local nor the global path ever gains access to the full feature representation from the preceding layer. In contrast, our S\textsuperscript{3}F-Net is a macro-level, \textbf{asymmetric two-tower architecture}, providing the \textbf{entire input signal} to both a deep spatial encoder and our shallow SpectraNet branch independently, allowing for an uncompromised analysis from each domain before a final fusion. This is a key conceptual advancement: this full-spectrum analysis, combined with a deliberately shallow summarizer design, is a more data-efficient architecture for injecting global context than stacking deep FFC blocks. Our approach also enhances fusion adaptability through the independent branches, making it possible to extend or modify the spectral encoder regardless of the structure of the spatial CNN branch, unlike FFC. 

\subsection{Feature Fusion Strategies}
As dual-branch and multi-modal architectures become more common, the strategy for fusing feature representations has become a critical area of research. The simplest method is \textbf{Concatenation Fusion}, where feature vectors from different branches are appended and passed to a dense layer to learn their combined representation~\cite{ngiam2011multimodal}. FiLM (Feature-wise Linear Modulation) uses a global context vector to generate affine transformation parameters that are applied to a primary feature map, allowing one branch to guide the other~\cite{perez2018film}.

A significant recent trend is the use of Transformer-based fusion mechanisms to explicitly capture both local and global context, a goal that aligns with the motivation for our S\textsuperscript{3}F-Net. For 2D medical image segmentation, architectures like MCBTNet integrate a CNN encoder for local features with a Transformer for global information, using bi-level routing attention to manage the fusion~\cite{zhang2025mcbtnet}. Similarly, in 3D remote sensing, dual-branch Transformers like HyperPointFormer use cross-attention to allow a geometric branch to be informed by a hyperspectral branch~\cite{rizaldy2025hyperpointformer}. While these hybrid CNN-Transformer models share our goal, our S\textsuperscript{3}F-Net explores a fundamentally different approach. Instead of employing a computationally intensive Transformer for the global branch, we propose the highly efficient, FFT-based \textbf{SpectraNet}, which achieves a global receptive field from its first layer with significantly lower overhead. A fusion mechanism that is more suitable for our task is \textbf{Bilinear Pooling}~\cite{lin2015bilinear}, a powerful technique that computes the outer product of the two feature vectors to model every pairwise interaction explicitly.

\section{Theoretical Framework} \label{sec:theory}


\subsection{The SpectralFilter Layer \& the Convolution Theorem}
Any signal in the spatial domain, such as an image, can be losslessly represented in the frequency domain. For a 2D image $f(x, y)$ of size $M \times N$, its Discrete Fourier Transform (DFT), $F(u, v)$, is defined as:
\begin{equation}
    F(u, v) = \sum_{x=0}^{M-1} \sum_{y=0}^{N-1} f(x, y) \cdot e^{-j 2\pi (\frac{ux}{M} + \frac{vy}{N})}
    \label{eq1}
\end{equation}
The output $F(u, v)$ is a complex-valued spectrum where each point represents the magnitude and phase of a sinusoidal plane wave of a specific frequency that constitutes the original image. The Fast Fourier Transform (FFT) computes the DFT with a lower computational complexity of $O(MN \log(MN))$.

The cornerstone of our learnable spectral filters is the Convolution Theorem, which provides a highly efficient alternative to spatial convolution. For two functions $f(x, y)$ (the image) and $h(x, y)$ (the convolutional filter), their spatial convolution $(f * h)(x, y)$ is defined as:
\begin{equation}
    (f * h)(x, y) = \sum_{m=0}^{M-1} \sum_{n=0}^{N-1} f(m, n) h(x-m, y-n)
\end{equation}
The Convolution Theorem states that this expensive operation is equivalent to a simple element-wise (Hadamard) product in the frequency domain:
\begin{equation}
    \mathcal{F}\{f * h\} = \mathcal{F}\{f\} \odot \mathcal{F}\{h\}
    \label{eq3}
\end{equation}
where $\mathcal{F}$ denotes the Fourier Transform. This allows us to compute the convolution by transforming both the image and the filter, multiplying them, and then transforming the result back to the spatial domain via the Inverse FFT (IFFT), $\mathcal{F}^{-1}$:
\begin{equation}
    f * h = \mathcal{F}^{-1}\{\mathcal{F}\{f\} \odot \mathcal{F}\{h\}\}
\end{equation}
Our proposed \textbf{SpectralFilter Layer} leverages this principle directly. Instead of learning a spatial filter $h$, we learn its frequency-domain representation $\mathcal{F}\{h\}$ as a set of complex-valued weights. This allows us to replace the spatial convolution with element-wise multiplication in the frequency domain.

\subsection{Instantaneous Global Receptive Field}
A key advantage of our spectral approach is its ability to achieve an instantaneous global receptive field. The receptive field of a neuron defines the specific region of the input image that influences its activation. Its size is therefore a critical measure of the model's ability to perceive and integrate contextual information. Whereas CNNs must gradually expand this field through depth, our spectral layer perceives the entire image context from the outset.

\textbf{Proof of Global Receptive Field:}
Let an input image be a 2D signal $f(x, y)$ and our learnable spectral filter be represented in the frequency domain as $H(u, v)$. The output of our \textbf{SpectralFilter Layer}, which we denote as the feature map $g(x, y)$, is computed via the Convolution Theorem as:
\begin{equation} \label{eq:output_g}
    g(x, y) = \mathcal{F}^{-1}\{F(u, v) \odot H(u, v)\}
\end{equation}
where $F(u, v) = \mathcal{F}\{f(x, y)\}$ and $\odot$ is the element-wise product. The process involves a forward DFT and an inverse DFT, both of which are global operations. By substituting the definition of the DFT (\ref{eq1}) into the inverse DFT and rearranging the summation terms, it can be formally shown that any output pixel $g(x, y)$ is a weighted sum over all input pixels $f(x', y')$:
\begin{flalign}
    & g(x, y) = \sum_{x'=0}^{M-1} \sum_{y'=0}^{N-1} f(x', y') \cdot \left[ \frac{1}{MN} \sum_{u=0}^{M-1} \sum_{v=0}^{N-1} H(u, v) \right. & \nonumber \\
    & \qquad \left. \cdot e^{j 2\pi (\frac{u(x-x')}{M} + \frac{v(y-y')}{N})} \right] &
\end{flalign}

Let the term in the brackets be $W(x,y,x',y')$. This term represents the weight of the contribution of the input pixel at $(x', y')$ to the output pixel at $(x, y)$. Crucially, this weight is non-zero for all pairs of $(x, y)$ and $(x', y')$. This equation demonstrates that the value of any given output pixel $g(x, y)$ is a weighted sum over \textit{every single pixel} $f(x', y')$ in the input image. Therefore, the receptive field of a single \textbf{SpectralFilter Layer} is the full size of the input image, $M \times N$.

On the other hand, the receptive field (RF) of a neuron in a CNN can be calculated iteratively. For a stack of convolutional layers, the RF of layer $l$ is given by:
\begin{equation}
    RF_l = RF_{l-1} + (k_l - 1) \times S_{l-1}
    \label{eq7}
\end{equation}
where $k_l$ is the kernel size of layer $l$, and $S_{l-1}$ is the product of all preceding strides. Even with a deep architecture like VGG16 operating on a $224 \times 224$ image, we can show using (\ref{eq7}) that the receptive field after all the convolutional layers is only $181 \times 181$. This neuron cannot see the entire $224 \times 224$ input image; it is blind to the corners and edges. To achieve a global view, such a network must rely on the final, destructive global pooling layer. In contrast, a single \textbf{SpectralFilter Layer}, by the proof above, has an instantaneous receptive field of $\textbf{224} \times \textbf{224}$ (or more depending on the input dimensions), demonstrating its efficiency at capturing global context.

Our proposed architecture is grounded on these theoretical principles. Since the point-wise multiplication operation needed is computationally efficient, and the global context is instantly available, we design a shallow spectral encoder with only 1 or 2 SpectralFilter layers across two variations. To distill that global information, we use depthwise separable convolution layers followed by a lightweight \textit{funnel} of dense layers to produce a considerably smaller feature vector. This compact feature vector also enables the use of Bi-linear Fusion, where an outer product of the two branch features is required, an operation that would significantly increase the number of parameters had the spectral feature vector been larger. A detailed breakdown of the architecture and the fusion mechanisms is provided in Section~\ref{sec:methodology}.

\section{Methodology} \label{sec:methodology}


\subsection{Datasets}
To validate the robustness and generalizability of our proposed framework, we conduct experiments across four diverse and challenging public medical imaging datasets.
\begin{itemize}
    \item \textbf{HAM10000:} Human Against Machine with 10,000 Training Images~\cite{tschandl2018ham10000} is a multi-class (7 classes) dermoscopy dataset for skin lesion classification. For evaluation, we use the official test set from the ISIC 2018 Challenge~\cite{codella2019skin}. The severe class imbalance present in this dataset is highly representative of real-world clinical data.
    \item \textbf{BRISC2025:} The Brain Tumor MRI Dataset for Segmentation and Classification~\cite{brisc2025_dataset} contains over 6,000 T1-weighted brain MRI scans. We utilize the classification task, which is a balanced, 4-class problem (glioma, meningioma, pituitary tumor, and no tumor).
    \item \textbf{Chest X-Ray (Pneumonia):} This is a large dataset of over 5,000 chest radiographs for the binary classification of Normal vs. Pneumonia~\cite{kermany2018chestxray}. The high resolution of these images is ideal for frequency analysis.
    \item \textbf{BUSI:} The Breast Ultrasound Images dataset~\cite{al2020busi} contains approximately 800 grayscale ultrasound images for a 3-class classification task (normal, benign, and malignant). The smaller scale of this dataset makes it extremely difficult to train deep networks from scratch, making it well-suited to demonstrate the data efficiency of S\textsuperscript{3}F-Net.
\end{itemize}

\subsection{Overall S\textsuperscript{3}F-Net Architecture}
\begin{figure*}[h]
    \centering
    \includegraphics[width=0.75\textwidth]{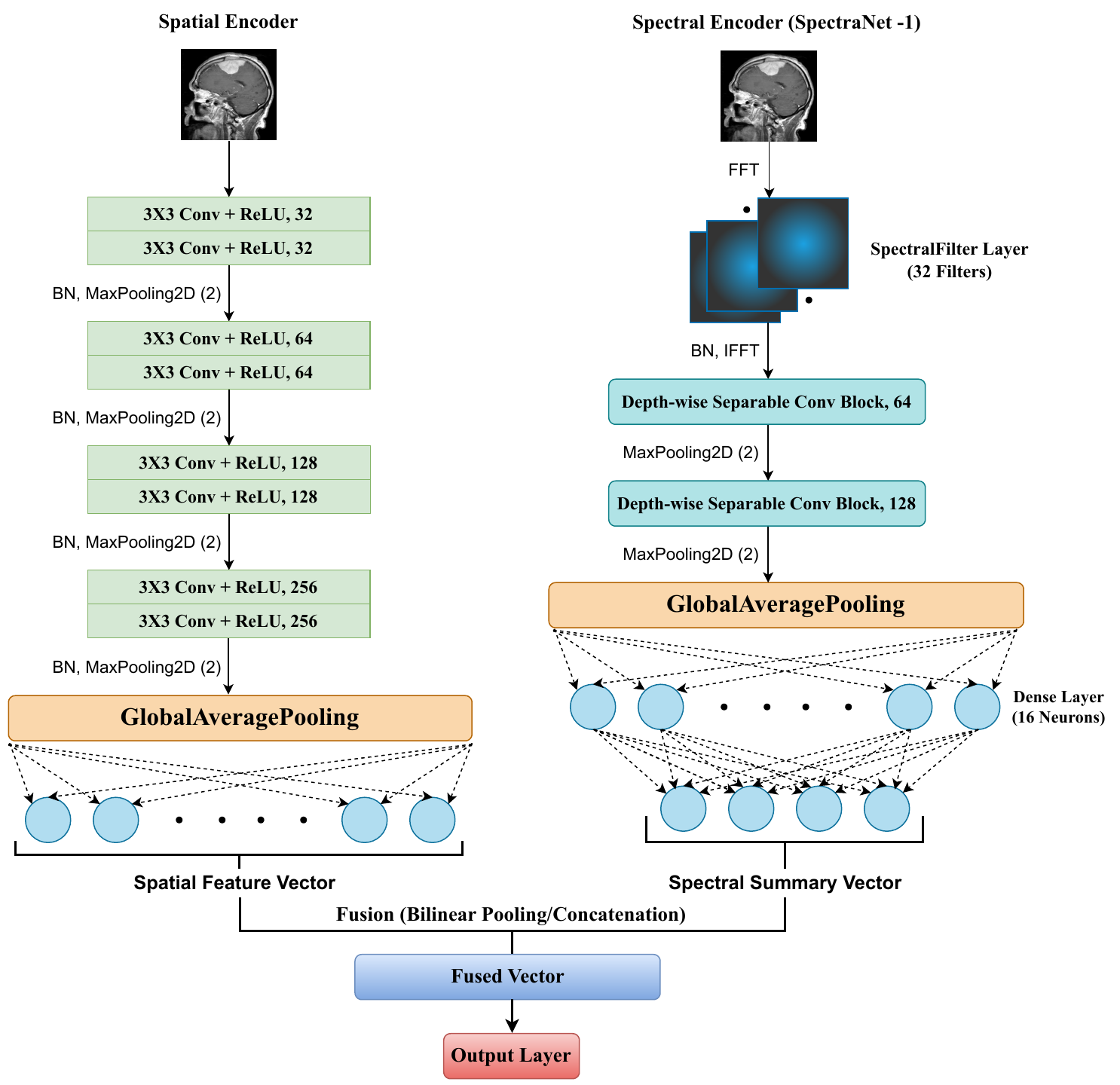}
    \caption{The architecture of the S\textsuperscript{3}F-Net, showcasing the asymmetric two-tower design with a deeper spatial encoder and a shallow SpectraNet branch (SpectraNet-1 in this figure), which uses a SpectralFilter layer and an efficient summarizer head to extract a global feature vector. The outputs of both branches are combined in a final fusion head for classification.}
    \label{fig:s3fnet_architecture}
\end{figure*}

The S\textsuperscript{3}F-Net employs an asymmetric two-tower design, as illustrated in Figure~\ref{fig:s3fnet_architecture}. The first tower is a deep spatial encoder responsible for learning a rich hierarchy of morphological features. The second tower is our novel, deliberately shallow \textbf{SpectraNet} branch, which efficiently extracts a low-dimensional summary of global, frequency-domain features. The feature vectors from these two branches are then combined using a fusion head, which acts as the input features for the final classifier and produces the final output.

\vspace{4 pt}
\subsubsection{The SpectralFilter Layer}
The foundational component of our spectral branch is the \textbf{SpectralFilter Layer}. This layer implements the Convolution Theorem (\ref{eq3}) directly, performing a global convolution via an efficient element-wise multiplication in the frequency domain.

The layer takes a 4D tensor (batch, height, width, channels) as input. Its internal weights, $W_{real}$ and $W_{imag}$, are learnable tensors defined in the frequency domain with a shape corresponding to the real-part FFT of the input. As standard deep learning optimizers operate on real-valued tensors, we represent each spectral filter $H(u,v)$ by parameterizing its real and imaginary components as two independent, trainable weight tensors, $W_{real}$ and $W_{imag}$. During the forward pass, these two tensors are combined to reconstruct the complex-valued filter $H(u, v) = W_{real} + j \cdot W_{imag}$ using a native complex number operation. This reconstructed filter is then multiplied with the complex-valued Fourier spectrum of the input. This approach allows the optimizer to update the real and imaginary parts of the spectral filter independently via standard backpropagation, while still correctly performing complex arithmetic in the forward pass. The forward pass, detailed in Algorithm~\ref{alg:spectralfilter}, is fully differentiable and can be integrated seamlessly into any deep learning framework.

\begin{algorithm}[h]
\caption{SpectralFilter Layer Forward Pass}
\label{alg:spectralfilter}
\begin{algorithmic}
\STATE \textbf{Input:} Image tensor $X \in \mathbb{R}^{B \times H \times W \times C_{in}}$
\STATE \textbf{Parameters:} $W_{real}, W_{imag} \in \mathbb{R}^{C_{in} \times C_{out} \times H \times (W/2+1)}$
\STATE 
\STATE $X_{perm} \leftarrow \text{Transpose}(X, \text{dims}=[0, 3, 1, 2])$
\STATE $X_{fft} \leftarrow \text{FFT}(X_{perm})$
\STATE $W_{fft} \leftarrow \text{Complex}(W_{real}, W_{imag})$
\STATE $Y_{fft} \leftarrow \text{Einsum}('bchw,cfhw \to bfhw', X_{fft}, W_{fft})$
\STATE $Y_{spatial} \leftarrow \text{IFFT}(Y_{fft})$
\STATE $Y_{perm} \leftarrow \text{Transpose}(Y_{spatial}, \text{dims}=[0, 2, 3, 1])$
\STATE \textbf{return} $Y_{perm} + \text{bias}$
\end{algorithmic}
\end{algorithm}
\vspace{4 pt}
\subsubsection{The SpectraNet Summarizer Branch}
We posit that since the SpectralFilter Layer has an instantaneous global receptive field, a deep spectral hierarchy is both unnecessary and computationally inefficient. Based on this, we designed the \textbf{SpectraNet} branch to act as a lightweight "summarizer" of global information.

We propose two shallow variants. The \textbf{SpectraNet-1} architecture (the spectral encoder shown in Figure~\ref{fig:s3fnet_architecture}), our primary design, utilizes a single SpectralFilter Layer at the full input resolution. Its output feature map is then passed through a lightweight head of Depthwise Separable Convolution blocks to summarize the spatial arrangement of the global frequency patterns. In contrast, the \textbf{SpectraNet-2} architecture stacks two SpectralFilter Layers (32 and 64 filters, respectively), which proved to be beneficial on the texture-dominant BUSI dataset.

A critical component of the summarizer head is the \textbf{Depthwise Separable Convolution block}~\cite{howard2017mobilenets, chollet2017xception}. The output of the SpectralFilter Layer is passed through these blocks to gradually downsample the feature map while efficiently learning to summarize the spatial arrangement of the global frequency patterns. A Depthwise Separable Convolution is a factorized version of a standard convolution that drastically reduces parameters and computation through two steps:

\begin{enumerate}
    \item \textbf{Depthwise Convolution:} A single filter is applied independently to each of the input channels. This step learns spatial patterns within each channel but does not combine information across channels.
    \item \textbf{Pointwise Convolution:} A simple $1 \times 1$ convolution with multiple filters is then used to project the output of the depthwise layer into the new channel space. This step learns linear combinations of the channels. 
\end{enumerate}

For a typical $3 \times 3$ kernel, this results in an 8-9X reduction in both parameters and computation, making it an ideal choice for a lightweight summarizer head. In our SpectraNet architecture, each of these blocks is followed by a \textbf{MaxPooling} layer, which summarizes the feature maps by reducing the heights and widths. The final stage of the SpectraNet is a \textbf{funnel} of two small dense layers. This was a key discovery of our research; this architectural bottleneck forces the model to distill the summarized features into a very small (4-dimensional) but powerful vector, which proved essential for effective fusion and preventing overfitting.
\vspace{4 pt}
\subsubsection{The Spatial Branch Baseline}
The spatial branch of the S\textsuperscript{3}F-Net shown in Figure~\ref{fig:s3fnet_architecture} serves two roles: it serves as the spatial encoder in our fusion model, and it acts as our primary experimental baseline. We employ a moderately deep, VGG-style architecture with four main blocks, each containing two $3 \times 3$ convolutional layers followed by Batch Normalization, a ReLU activation, MaxPooling, and Dropout. We deliberately chose this architecture over much deeper models like ResNet50 since our SpectraNet branch is capturing the global receptive field instantaneously, and a massively deep spatial network is thus less necessary.

\subsection{Fusion Strategies}
Let the spatial tower output be the vector $\mathbf{v}_{s} \in \mathbb{R}^{d_s}$ and the SpectraNet branch output be the vector $\mathbf{v}_{f} \in \mathbb{R}^{d_f}$ (where $d_f=4$). We investigated two primary fusion mechanisms:

\begin{enumerate}
    \item \textbf{Concatenation Fusion:} This is a simple and robust baseline fusion where the two vectors are concatenated to form a single, larger vector $\mathbf{z}_{cat}$.
    \begin{equation}
        \mathbf{z}_{cat} = [\mathbf{v}_{s} ; \mathbf{v}_{f}] \in \mathbb{R}^{d_s + d_f}
    \end{equation}
    This combined vector is then passed to a dense layer to learn the relationships between the two feature sets.

    \item \textbf{Bilinear Fusion:} To explicitly model the rich, pairwise interactions between the two feature domains, we employ Bilinear Pooling~\cite{lin2015bilinear}. This process creates a highly expressive feature vector by computing the outer product of the two branch outputs, proceeding in three steps:
    \begin{enumerate}[label=(\roman*)]
        \item First, the refined spatial vector $\mathbf{v}_{s} \in \mathbb{R}^{d_s}$ and the summarized spectral vector $\mathbf{v}_{f} \in \mathbb{R}^{d_f}$ are prepared.
        \item Next, their outer product is computed via matrix multiplication to form an interaction matrix $\mathbf{P}$:
        \begin{equation}
            \mathbf{P} = \mathbf{v}_{s} \mathbf{v}_{f}^T \in \mathbb{R}^{d_s \times d_f}
        \end{equation}
        Each element $P_{ij}$ of this matrix represents the multiplicative interaction between the $i$-th spatial feature and the $j$-th spectral feature.
        \item Finally, the matrix $\mathbf{P}$ is flattened (vectorized) to form the final fused feature vector, $\mathbf{z}_{bilinear}$:
        \begin{equation}
            \mathbf{z}_{bilinear} = \text{vec}(\mathbf{P}) \in \mathbb{R}^{d_s d_f}
        \end{equation}
    \end{enumerate}
    This vector captures the second-order statistics between the two domains, modeling more complex relationships.
\end{enumerate}


\section{Experimental Setup}
We evaluate our models on four diverse, publicly available medical imaging datasets: HAM10000~\cite{tschandl2018ham10000}, BRISC2025~\cite{brisc2025_dataset}, Chest X-Ray (Pneumonia)~\cite{kermany2018chestxray}, and BUSI~\cite{al2020busi}. We used the predefined publicly available Training and Test sets for all datasets. All models were implemented in TensorFlow and trained from scratch on an NVIDIA A100 GPU. We used the Adam optimizer with an initial learning rate of $3 \times 10^{-4}$ and a ReduceLROnPlateau callback. Due to the significant class imbalance present in some datasets, we employed balanced class weights during training and selected the best model based on the validation set's Weighted F1-Score, a metric robust to class imbalance, while balancing overall performance. For our final analysis, we report four key metrics on the unseen test set: Accuracy, Weighted F1-Score, Area Under the Receiver Operating Characteristic Curve (AUC-ROC), and the Matthews Correlation Coefficient (MCC).

\section{Results and Discussion} \label{sec:results}


\subsection{Ablation Study on Spectral Filter Count}

While most \textbf{SpectraNet} hyperparameters were determined through extensive tuning on the HAM10000 and Chest X-Ray datasets, the number of filters in our initial \textbf{SpectralFilter Layer} warranted a specific ablation study to balance performance and efficiency. We tested this on our best-performing S\textsuperscript{3}F-Net (Concatenation) for the Chest X-Ray dataset.

The results, shown in Table~\ref{tab:filter_ablation}, depict that an initial layer with \textbf{32 filters} achieves a clear peak in performance, reaching 93.11\% accuracy. Increasing the filter count to 64 degraded performance, likely due to overfitting. Using 16 filters reduces the parameter count, but performance drops significantly. This analysis shows that 32 is the optimal filter count for this architecture, and this value was used for all reported experiments.

\begin{table}[h]
\centering
\caption{Ablation study on the number of initial spectral filters, evaluated on the Chest X-Ray test set.}
\label{tab:filter_ablation}
\begin{tabular}{l|c|ccc}
\toprule
\textbf{Filter Count} & \textbf{\# Params} & \textbf{Accuracy} & \textbf{F1-score} & \textbf{AUC-ROC} \\
\midrule
16 Filters & \textbf{4.48M} & 0.9215 & 0.9384 & 0.9721 \\
\textbf{32 Filters} & 7.64M & \textbf{0.9311} & \textbf{0.9453} & \textbf{0.9749} \\
64 Filters & 13.95M & 0.9183 & 0.9344 & 0.9673 \\
\bottomrule
\end{tabular}
\end{table}

\subsection{Ablation Study: Standalone Branch Performance}
To understand the individual contributions of each domain, we conducted an ablation study by evaluating the Spatial-Only and Spectral-Only models in isolation.

A key finding emerged from this ablation study, as shown in Table~\ref{tab:main_results}. Although the Spatial Baseline performed better across most datasets, as expected, the SpectraNet models demonstrated remarkable and surprising performance on the Chest X-Ray dataset. The shallow \textbf{SpectraNet-1} model significantly outperformed the baseline CNN in terms of all metrics, as did \textbf{SpectraNet-2}. SpectraNet-1, with only 1 SpectralFilter Layer, outperformed the much deeper baseline CNN by a margin of over \textbf{3\%} in terms of accuracy. We hypothesize this is due to a combination of two factors: the pathology of pneumonia, which often manifests as diffuse textural changes, is inherently well-suited to a global frequency-domain analysis; and the high-resolution nature of the X-ray images provides a high-fidelity signal for the FFT-based \textit{SpectralFilter Layer}. This result supports our hypothesis and provides the core motivation for a fusion model, showing that neither domain is universally superior and that both provide unique, valuable information.

\subsection{Baseline Comparison \& Analyzing Fusion Strategies}
We now present the core results of our study in Table~\ref{tab:main_results}. This table compares our two primary S\textsuperscript{3}F-Net fusion architectures against the strong Spatial Baseline across all four datasets. S\textsuperscript{3}F-Net (Concat.) refers to the Concatenation Fusion, whereas S\textsuperscript{3}F-Net (Bilinear) refers to the Bilinear Pooling Fusion. For each fusion method, we tested both the SpectraNet-1 (SN1) and SpectraNet-2 (SN2) variants. We trained and tested each model three times, and took the average and standard deviation. The best models and the baselines were evaluated five times each to ensure reproducibility.

\begin{table*}[t]
\centering
\caption{Main Performance Comparison on Test Sets Across All Datasets. Results are reported as Mean ± Standard Deviation over multiple independent runs. Asterisks denote a statistically significant improvement over the Spatial Baseline for that metric (*p $<$ 0.05, **p $<$ 0.01).}
\label{tab:main_results}
\begin{tabular}{llcccc}
\toprule
\textbf{Dataset} & \textbf{Model} & \textbf{Accuracy} & \textbf{F1-Score (Weighted)} & \textbf{AUC-ROC} & \textbf{MCC} \\ 
\midrule
\multirow{7}{*}{\textbf{HAM10000}} & Spatial Baseline & 0.6671 ± 0.0180 & 0.6889 ± 0.0155 & 0.8884 ± 0.0066 & 0.4930 ± 0.0025 \\ \addlinespace
 & SpectraNet-1 & 0.6310 ± 0.0212 & 0.6408 ± 0.0243 & 0.8324 ± 0.0155 & 0.4074 ± 0.0281 \\
 & SpectraNet-2 & 0.6448 ± 0.0222 & 0.6402 ± 0.0252 & 0.8529 ± 0.0133 & 0.4054 ± 0.0252 \\ \addlinespace
 & S\textsuperscript{3}F-Net (Concat., SN1) & \textbf{0.7011 ± 0.0096*} & 0.6974 ± 0.0093 & 0.8820 ± 0.0013 & 0.4924 ± 0.0077 \\
 & S\textsuperscript{3}F-Net (Concat., SN2) & 0.6931 ± 0.0112 & 0.6841 ± 0.0121 & \textbf{0.8905 ± 0.0027} & 0.4765 ± 0.0098 \\
 & \textbf{S\textsuperscript{3}F-Net (Bilinear, SN1)} & 0.6931 ± 0.0000* & \textbf{0.7038 ± 0.0013*} & 0.8892 ± 0.0066 & \textbf{0.5136 ± 0.0069*} \\
 & S\textsuperscript{3}F-Net (Bilinear, SN2) & 0.6071 ± 0.0220 & 0.6311 ± 0.0241 & 0.8409 ± 0.0163 & 0.4236 ± 0.0310 \\ 
\midrule
\multirow{7}{*}{\textbf{BRISC2025}} & Spatial Baseline & 0.9827 ± 0.0007 & 0.9827 ± 0.0006 & 0.9982 ± 0.0007 & 0.9764 ± 0.0010 \\ \addlinespace
 & SpectraNet-1 & 0.9446 ± 0.0064 & 0.9442 ± 0.0063 & 0.9912 ± 0.0018 & 0.9244 ± 0.0072 \\
 & SpectraNet-2 & 0.9525 ± 0.0055 & 0.9524 ± 0.0054 & 0.9935 ± 0.0015 & 0.9352 ± 0.0062 \\ \addlinespace
 & S\textsuperscript{3}F-Net (Concat., SN1) & 0.9772 ± 0.0021 & 0.9772 ± 0.0021 & 0.9976 ± 0.0008 & 0.9689 ± 0.0031 \\
 & S\textsuperscript{3}F-Net (Concat., SN2) & 0.9703 ± 0.0032 & 0.9703 ± 0.0031 & 0.9961 ± 0.0011 & 0.9595 ± 0.0040 \\
 & \textbf{S\textsuperscript{3}F-Net (Bilinear, SN1)} & \textbf{0.9876 ± 0.0007*} & \textbf{0.9876 ± 0.0007*} & 0.9982 ± 0.0001 & \textbf{0.9831 ± 0.0010*} \\
 & S\textsuperscript{3}F-Net (Bilinear, SN2) & 0.9792 ± 0.0018 & 0.9792 ± 0.0019 & \textbf{0.9985 ± 0.0004} & 0.9716 ± 0.0025 \\ 
\midrule
\multirow{7}{*}{\textbf{Chest X-Ray}} & Spatial Baseline & 0.8798 ± 0.0204 & 0.9054 ± 0.0117 & 0.9433 ± 0.0127 & 0.7440 ± 0.0489 \\ \addlinespace
 & SpectraNet-1 & 0.9103 ± 0.0110 & 0.9314 ± 0.0088 & 0.9659 ± 0.0052 & 0.8090 ± 0.0210 \\
 & SpectraNet-2 & 0.9022 ± 0.0122 & 0.9259 ± 0.0096 & 0.9573 ± 0.0065 & 0.7928 ± 0.0243 \\ \addlinespace
 & \textbf{S\textsuperscript{3}F-Net (Concat., SN1)} & \textbf{0.9311 ± 0.0125**} & \textbf{0.9453 ± 0.0097**} & 0.9749 ± 0.0067 & \textbf{0.8524 ± 0.0270**} \\
 & S\textsuperscript{3}F-Net (Concat., SN2) & 0.9271 ± 0.0079 & 0.9422 ± 0.0052 & \textbf{0.9753 ± 0.0040**} & 0.8443 ± 0.0180 \\
 & S\textsuperscript{3}F-Net (Bilinear, SN1) & 0.8654 ± 0.0181 & 0.9007 ± 0.0192 & 0.8959 ± 0.0152 & 0.7164 ± 0.0355 \\
 & S\textsuperscript{3}F-Net (Bilinear, SN2) & 0.8750 ± 0.0170 & 0.9071 ± 0.0177 & 0.9286 ± 0.0131 & 0.7363 ± 0.0311 \\ 
\midrule
\multirow{7}{*}{\textbf{BUSI}} & Spatial Baseline & 0.8430 ± 0.0046 & 0.8437 ± 0.0059 & 0.9505 ± 0.0059 & 0.7345 ± 0.0131 \\ \addlinespace
 & SpectraNet-1 & 0.8269 ± 0.0081 & 0.8264 ± 0.0091 & 0.9099 ± 0.0112 & 0.7029 ± 0.0183 \\
 & SpectraNet-2 & 0.8205 ± 0.0095 & 0.8175 ± 0.0103 & 0.9294 ± 0.0093 & 0.6905 ± 0.0210 \\ \addlinespace
 & S\textsuperscript{3}F-Net (Concat., SN1) & 0.8526 ± 0.0031 & 0.8496 ± 0.0040 & 0.9565 ± 0.0059 & 0.7440 ± 0.0110 \\
 & \textbf{S\textsuperscript{3}F-Net (Concat., SN2)} & \textbf{0.8782 ± 0.0000**} & \textbf{0.8771 ± 0.0004**} & \textbf{0.9601 ± 0.0064*} & \textbf{0.7911 ± 0.0020**} \\
 & S\textsuperscript{3}F-Net (Bilinear, SN1) & 0.7949 ± 0.0120 & 0.7910 ± 0.0131 & 0.9000 ± 0.0142 & 0.6601 ± 0.0251 \\
 & S\textsuperscript{3}F-Net (Bilinear, SN2) & 0.8077 ± 0.0110 & 0.8031 ± 0.0111 & 0.8778 ± 0.0189 & 0.6795 ± 0.0227 \\
\bottomrule
\multicolumn{6}{l}{\footnotesize The best performing model and the best metrics for each dataset are highlighted in \textbf{bold}.}
\end{tabular}
\end{table*}

The results presented in Table~\ref{tab:main_results} lead to several key conclusions. Firstly, across all four datasets and all primary metrics, a version of our \textbf{S\textsuperscript{3}F-Net consistently and significantly outperforms the strong Spatial Baseline}. This validates our core hypothesis that fusing separate branches trained on spatial and spectral features provides a more powerful and generalizable representation for medical image analysis, providing robust evidence for our model's superior performance, as it achieves simultaneous improvement across four distinct and complementary metrics: Accuracy (overall correctness), F1-Score (balanced precision-recall), AUC-ROC (diagnostic separability), and MCC (correlation on imbalanced classes).

Secondly, the results reveal that the \textbf{optimal fusion strategy is highly task-dependent}. For the complex, multi-featured classification tasks of HAM10000 (dermoscopy) and BRISC2025 (MRI), the \textbf{Bilinear Fusion} mechanism was the best performer. Its ability to model the pairwise interactions between the spatial and spectral domains proved essential for distinguishing between visually similar classes. For instance, on BRISC2025, the S\textsuperscript{3}F-Net (Bilinear, SN1) achieves a state-of-the-art competitive accuracy of \textbf{98.76\%}, surpassing the already strong Spatial Baseline by a notable margin. 

This advantage is even more apparent on the HAM10000 dataset. S\textsuperscript{3}F-Net achieves the highest W. F1-Score (\textbf{0.7038}) and MCC (\textbf{0.5136}), surpassing all other variants, including the baseline. This indicates that the Bilinear Fusion is not just improving performance on the majority class, but is learning a more nuanced decision boundary that leads to better-balanced, class-wise performance. It shows that by explicitly modeling the interaction between a specific morphological feature and a specific textural feature, the Bilinear model can make more confident and accurate predictions on the difficult, rare classes, which is critical for a diagnostically useful model.

Conversely, for the datasets that are more dominated by global textural features, namely Chest X-Ray and BUSI, the simpler \textbf{Concatenation Fusion} proved to be more effective. On the Chest X-Ray dataset, the S\textsuperscript{3}F-Net (Concat., SN1) achieved a remarkable \textbf{93.11\%} accuracy, a greater than \textbf{5\%} absolute improvement over the Spatial Baseline. Similarly, on the data-scarce BUSI dataset, the S\textsuperscript{3}F-Net (Concat., SN2) achieved the highest accuracy of \textbf{87.82\%}. We hypothesize that for these tasks, the clean separation of features provided by concatenation is a more robust strategy, preventing the model from overfitting on spurious feature interactions that the more powerful Bilinear Fusion might find. The success of the SpectraNet-2 on BUSI also suggests that texture-dominant modalities may benefit from a slightly deeper spectral analysis.

To ensure the performance gains of our S\textsuperscript{3}F-Net weren't incidental, we ran a full statistical significance analysis. We performed an independent two-sample t-test (Welch's t-test) comparing our best S\textsuperscript{3}F-Net variant against the Spatial Baseline across five independent runs for each dataset. As the asterisks in Table~\ref{tab:main_results} denote, the improvements in key metrics like Accuracy and F1-Score were found to be \textbf{statistically significant (p $<$ 0.05)} across all four domains. The significance was particularly important for the Chest X-Ray dataset, our best case, where the S\textsuperscript{3}F-Net's accuracy improvement yielded a p-value of just \textbf{0.0013}. This crucial finding allows us to reject the null hypothesis and conclude that our proposed spatial-spectral summarizer fusion network provides a consistent performance advantage over a traditional CNN.

\subsection{Comparison with State-of-the-Art Models}
We compare the best-performing variants of S\textsuperscript{3}F-Net against a suite of well-established deep learning architectures. On the BRISC2025 dataset, we compare against standard ImageNet pre-trained models to demonstrate SOTA-competitiveness, whereas on the Chest X-Ray dataset, we compare against models trained from scratch to highlight data efficiency.

\subsubsection{Performance on BRISC2025}
The BRISC2025 dataset, with its high-quality images, serves as an excellent benchmark for top-end performance. As shown in Table~\ref{tab:sota_brisc}, our S\textsuperscript{3}F-Net with Bilinear Fusion, despite being trained from scratch, achieves a remarkable accuracy of \textbf{98.76\%}. This result surpasses deeper and more parameter-heavy models like VGG, ResNet, DenseNet, and Inception variants. Our model also significantly outperforms these complex architectures in all other metrics. Although EfficientNetB0 still holds the top position, our S\textsuperscript{3}F-Net demonstrates that a novel, data-efficient architecture can achieve performance within the elite tier without relying on large-scale pre-training. Also, our model achieves this result with the lowest number of parameters (\textbf{3.44M}), making it suitable to employ even in edge devices.

\begin{table}[h]
\centering
\caption{S\textsuperscript{3}F-Net Performance on BRISC2025 Compared to Standard SOTA Models as Provided in the Original Paper~\cite{brisc2025_dataset}.}
\label{tab:sota_brisc}
\begin{tabular}{l|rcccc}
\toprule
\textbf{Model} & \textbf{\# Params} & \textbf{Acc.} & \textbf{F1} & \textbf{Precision} & \textbf{Recall} \\
\midrule
VGG16 & 138.4M & 0.9692 & 0.9693 & 0.9706 & 0.9693 \\
VGG19 & 143.7M & 0.9629 & 0.9630 & 0.9639 & 0.9630 \\
ResNet50 & 25.6M & 0.9820 & 0.9820 & 0.9823 & 0.9820 \\
ResNet101 & 44.5M & 0.9809 & 0.9810 & 0.9815 & 0.9810 \\
DenseNet169 & 14.3M & 0.5093 & 0.5710 & 0.6404 & 0.5710 \\
InceptionV3 & 23.8M & 0.6198 & 0.6487 & 0.7225 & 0.6487 \\
EfficientNetB0 & 5.3M & 0.9920 & 0.9920 & 0.9920 & 0.9920 \\
MobileNetV3 & 5.4M & 0.9442 & 0.9447 & 0.9475 & 0.9447 \\
\midrule
\textbf{S\textsuperscript{3}F-Net} & \textbf{3.44M} & \textbf{0.9876} & \textbf{0.9876} & \textbf{0.9871} & \textbf{0.9891} \\
\bottomrule
\end{tabular}
\end{table}

\subsubsection{Performance on Chest X-Ray Dataset}
The Chest X-Ray dataset provides an ideal environment to test our central hypothesis regarding data efficiency. We compare our best-performing S\textsuperscript{3}F-Net with Concatenation Fusion against a range of standard CNNs trained entirely from scratch, using benchmark results from a recent study by Alanazi et al.~\cite{alanazi2024chestxray_fromscratch}.

\begin{table}[h]
\centering
\caption{S\textsuperscript{3}F-Net Performance on Chest X-Ray Pneumonia Compared to SOTA Models Trained From Scratch. Data for other models sourced from~\cite{alanazi2024chestxray_fromscratch},~\cite{Alshanketi2025Pneumonia}.}
\label{tab:sota_xray}
\begin{tabularx}{\columnwidth}{>{\raggedright\arraybackslash}X|rcccc} 
\toprule
\textbf{Model} & \textbf{\# Params} & \textbf{Acc.} & \textbf{F1} & \textbf{Precision} & \textbf{Recall} \\
\midrule
ResNet18 & 11.7M & 0.8960 & 0.9180 & 0.9010 & 0.9360 \\
ResNet50 & 25.6M & 0.8300 & 0.8610 & 0.8840 & 0.8380 \\
VGG16 & 138.4M & 0.9100 & 0.9310 & 0.8990 & 0.9640 \\
MobileNetV2 & 3.4M & 0.8000 & 0.8300 & 0.8820 & 0.7850 \\
InceptionV3 & 23.8M & 0.9120 & 0.9330 & 0.8970 & 0.9720 \\
DenseNet169 & 14.3M & 0.9120 & 0.9330 & 0.8920 & 0.9769 \\
\midrule
\textbf{S\textsuperscript{3}F-Net} & \textbf{7.64M} & \textbf{0.9311} & \textbf{0.9453} & \textbf{0.9381} & \textbf{0.9530} \\
\bottomrule
\end{tabularx}
\end{table}

The results in Table~\ref{tab:sota_xray} are striking. Our S\textsuperscript{3}F-Net not only achieves the highest accuracy (\textbf{93.11\%}) and the F1 score (\textbf{0.9453}) by a significant margin, but it does so with only \textbf{7.64M} parameters. It surpasses much deeper models like ResNet50 and DenseNet169, which have 2-3 times the parameter count. S\textsuperscript{3}F-Net also demonstrates a much more balanced Precision-Recall trade-off with a precision of \textbf{93.81\%} and recall of \textbf{95.3\%}.


The results discussed validate that our dual-domain approach, which extracts global, frequency-domain features via the shallow SpectraNet branch, allows the model to learn a more robust representation from limited training data than purely spatial architectures. This inherent architectural advantage makes the S\textsuperscript{3}F-Net a powerful and efficient framework for tasks where large-scale pre-training is not feasible.


\subsection{Explainability Analysis}

To understand \textit{why} our S\textsuperscript{3}F-Net framework is so effective, we perform an explainability analysis on the best-performing Chest X-ray model to quantify the influence of each branch on the final decision. Standard metrics like the L2 Norm are biased by vector dimensionality. To overcome this, we develop a novel, \textbf{balanced contribution score ($C_v$)} for a feature vector $\mathbf{v} \in \mathbb{R}^{d}$, defined as the geometric mean of its total magnitude and its per-feature magnitude:
\begin{equation}
    C_v = \sqrt{ ||\mathbf{v}||_2 \cdot \frac{||\mathbf{v}||_2}{d} } = \frac{||\mathbf{v}||_2}{\sqrt{d}}
\end{equation}
This score fairly compares the influence of vectors with disparate dimensionalities by rewarding both total signal strength and intensity per feature.

\begin{figure}[h]
    \centering
    \includegraphics[width=\columnwidth]{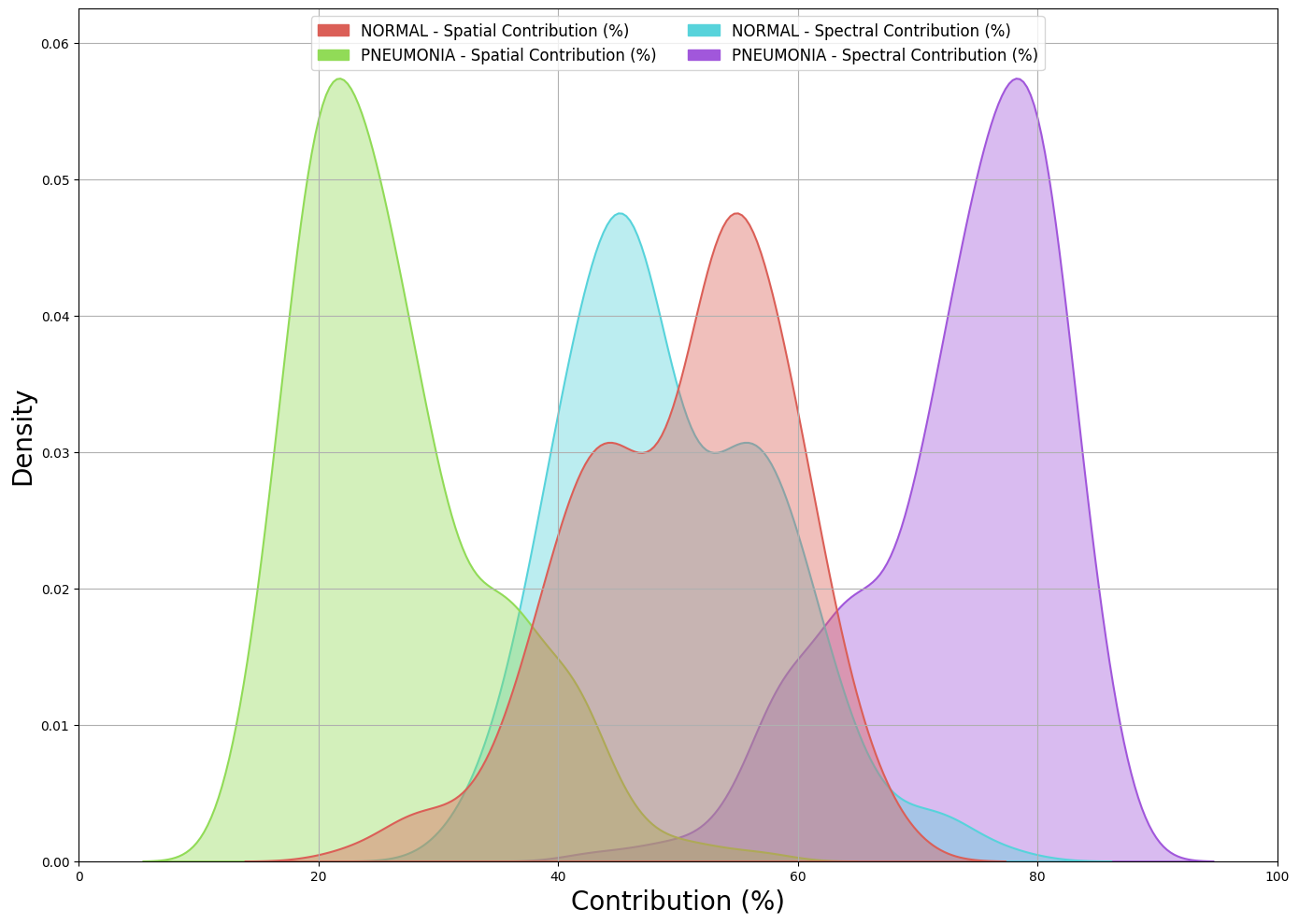}
    \caption{Combined distribution of the contribution scores for the Spatial and Spectral branches on the Chest X-Ray dataset, separated by true class. It shows different distributions for different classes.}
    \label{fig:contribution_plot}
\end{figure}

By extracting the refined feature vectors from each tower before fusion, we analyzed their relative contribution scores on the test set. The results, shown in Figure~\ref{fig:contribution_plot}, reveal an interesting phenomenon: the S\textsuperscript{3}F-Net dynamically alters its reliance on each branch based on the input's class.

For \textbf{Normal} X-rays, the contributions are almost perfectly balanced (50.52\% Spatial vs. 49.48\% Spectral), suggesting the model requires a consensus from both experts for a "healthy" diagnosis. In contrast, for X-rays exhibiting \textbf{Pneumonia}, the model overwhelmingly relies on the Spectral branch, with an average spectral contribution of \textbf{73.64\%}. This provides powerful quantitative evidence for our hypothesis: the diffuse, textural nature of pneumonia is more effectively captured by the global, frequency-domain analysis of our SpectraNet.

\begin{figure*}[h]
    \centering
    \begin{subfigure}[b]{0.7\columnwidth}
        \includegraphics[width=\textwidth]{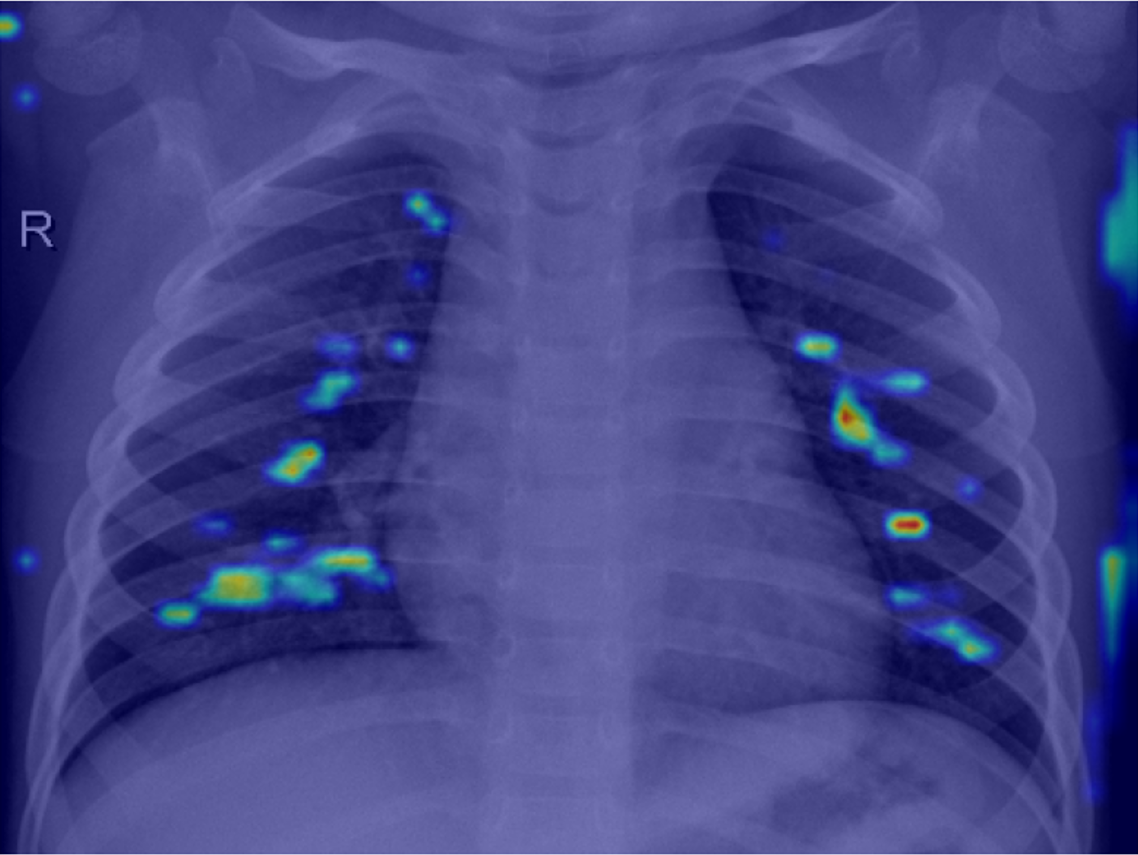}
        \caption{S\textsuperscript{3}F-Net focus for a NORMAL case.}
        \label{fig:gradcam_normal}
    \end{subfigure}
    \qquad 
    \begin{subfigure}[b]{0.7\columnwidth}
        \includegraphics[width=\textwidth]{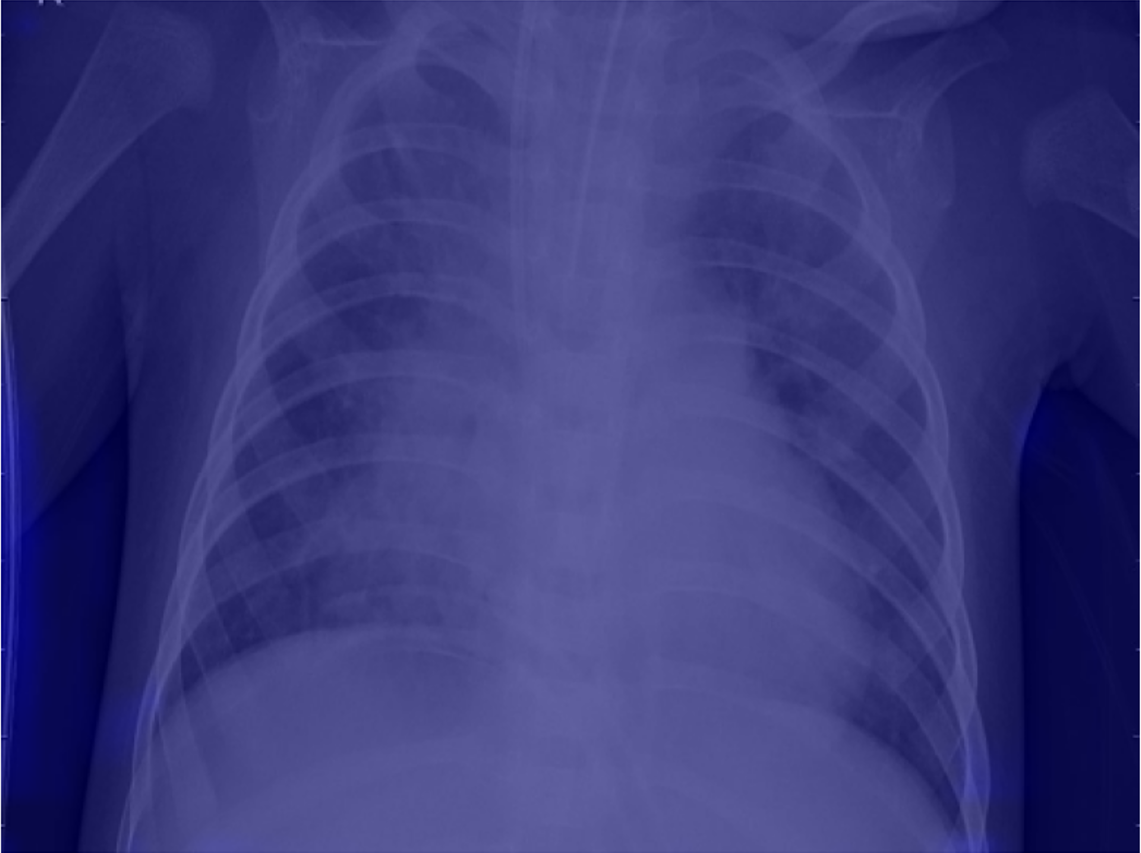}
        \caption{S\textsuperscript{3}F-Net focus for a PNEUMONIA case.}
        \label{fig:gradcam_pneumonia}
    \end{subfigure}
    \caption{Grad-CAM saliency analysis. The active spatial focus on anatomical landmarks in (a) contrasts with the inactive map in (b), highlighting the model's reliance on its spectral branch to detect the diffuse texture of pneumonia.}
    \label{fig:gradcam_analysis}
\end{figure*}

It should be noted that $C_v$ is not a strict explainability metric. It mainly reflects the relative variation in branch activation strength. To provide a more direct interpretation of the model's focus, we supplemented our analysis with Gradient-weighted Class Activation Mapping (Grad-CAM)~\cite{selvaraju2017grad}. The resulting saliency maps (Fig.~\ref{fig:gradcam_analysis}) reveal a sophisticated reasoning process.

For a correctly classified \textbf{NORMAL} case (Fig.~\ref{fig:gradcam_normal}), the Grad-CAM shows that the model's attention is distributed across prominent, high-contrast anatomical structures, particularly the ribs and the cardiomediastinal silhouette, essentially confirming the absence of anomalies. In contrast, for a correctly classified \textbf{PNEUMONIA} case (Fig.~\ref{fig:gradcam_pneumonia}), the spatial activation map is almost entirely \textit{cold}, showing no significant activation within the lung fields. This result is a powerful visual validation of our dual-branch hypothesis. It confirms that the classification is not being driven by any specific local feature, but rather by the global, textural information captured by the SpectraNet branch, which our contribution analysis showed was dominant for this pathology.

This analysis validates that our fusion approach is not merely combining signals, but is facilitating a dynamic strategy that enables the multi-modal model to learn to weigh different input modalities based on the pathology it observes.

\section{Conclusion} \label{sec:conclusion}

In this work, we addressed two fundamental limitations of standard CNNs in medical imaging: their inherent locality and their dependence on a single modality of signal representation. We introduced the \textbf{S\textsuperscript{3}F-Net}, a novel dual-branch framework that learns from both spatial and spectral domains. This successful integration of two complementary domains through the shallow \textbf{SpectraNet} branch leads our model to outperform its strong spatial-only baseline across four diverse medical datasets consistently and significantly, validating our core hypothesis. We also showed that the optimal fusion mechanism is demonstrably \textbf{task-dependent} in this case. 

While our results are highly competitive, some limitations present opportunities for future work. The performance on the HAM10000 dataset, while strong, indicates that further research into class-balancing techniques for dual-branch models is warranted. Also, the scalability of our framework to 3D imaging modalities like volumetric CT or MRI remains to be explored. Another limitation is the memory requirements for SpectraNet-2 variants. Storing its full-resolution frequency-domain weights could pose a challenge for deployment on resource-constrained edge devices.

Despite these, through explicitly and efficiently engineering a branch to capture global context, S\textsuperscript{3}F-Net achieves a level of data efficiency that is crucial in the often data-scarce medical domain. The powerful S\textsuperscript{3}F-Net encoder can be directly adapted as a backbone for segmentation and other prediction tasks where the interplay of local and global context is critical, opening several promising avenues for future research.



\section*{Acknowledgement}

The authors gratefully acknowledge the vital institutional support provided by BRAC University and the Department of Electrical and Electronic Engineering (EEE) at Bangladesh University of Engineering and Technology (BUET).

\end{document}